\documentclass{emulateapj}

\begin{document}

\title{Survey of Nearby FGK Stars at 160 $\micron$ with Spitzer}

\author{Angelle Tanner\altaffilmark{1,2}, 
Charles Beichman\altaffilmark{3}, 
Geoff Bryden\altaffilmark{2},
Carey Lisse\altaffilmark{4},
Samantha Lawler\altaffilmark{5}
}

\altaffiltext{1}{Department of Physics and Astronomy, Georgia State University, Atlanta, GA 30302-4106; angelle.tanner@gmail.com}
\altaffiltext{2}{Jet Propulsion Lab, 4800 Oak Grove Dr., Pasadena, CA 91109}
\altaffiltext{3}{NASA Exoplanet Science Institute, Caltech, 770 S. Wilson Ave., Pasadena, CA 91125 }
\altaffiltext{4}{Johns Hopkins University - Applied Physics Laboratory , SD/SRE, MP3-W155, 7707 Montpelier Road Laurel, MD 20723}
\altaffiltext{5}{Astronomy Department, Van Vleck Observatory, Wesleyan University, Middletown, CT 06459}

\begin{abstract}

The Spitzer Space Telescope has advanced debris disk science tremendously with 
a wealth of information on  debris disks around nearby A, F, G, K and M stars at 24 and 70 $\micron$ with the MIPS photometer and at 8-34 $\micron$ with IRS. Here we present 160 $\micron$ observations of a small sub-set of these stars. At this  wavelength, the stellar photospheric emission is negligible and any detected emission corresponds to cold dust in extended Kuiper belt analogs. However, the Spitzer 160 $\micron$ observations are limited in sensitivity by the large beam size which results in significant ``noise'' due to cirrus and extragalactic confusion. In addition, the 160 $\micron$ measurements suffer from the added complication of a light leak next to the star's position whose flux is proportional to the near-infrared flux of the star. We are able to remove the contamination from the leak and report  160 $\micron$ measurements or upper limits  for 24  stars. Three stars (HD 10647, HD 207129, and HD 115617) have excesses at 160 $\micron$ that we  use to constrain the properties of the debris disks around them. A more detailed model of the spectral energy distribution of HD 10647 reveals that the 70 and 160 $\micron$ emission could be due to small water ice particles at a distance of 100 AU consistent with Hubble Space Telescope optical imaging of circumstellar material in the system. 

\end{abstract}


\section{Introduction}

Surveys of nearby main-sequence stars with the {\it Spitzer Space Telescope} have produced a new sample of debris disks which will be  studied for years to come.  Since Poynting-Robinson (P-R) drag, radiation pressure and wind 
drag\footnote{Some studies suggests that stellar wind drag could be the dominant force in removing grains from debris disks around M and K dwarfs (Chen et al. 2006; Plavchan et al. 2009).} remove dust from the disk, the dust in debris disk systems must be replenished, presumably through collisions between larger bodies, possibly  remnant planetesimals that were the building blocks of  full-fledged planets.  At 70 $\micron$ {\it Spitzer} MIPS photometry is sensitive to dust emission which is more than 5-10 times more luminous than that seen in our own solar system  with  excesses analogous to emission from our Kuiper belt  occurring in $\sim$15\% of main sequence stars (Bryden et al. 2006). At 8-14  $\micron$ {\it Spitzer} is sensitive to dust emission which is more than 1,000 times more luminous than that seen in our own solar system  and excesses analogous to emission from our own asteroid the Kuiper belt are found to occur in roughly 1-2\% of main sequence stars (Beichman et al. 2006). This paper focuses on the detection of  excesses at 160 $\micron$ thus probing the properties of Kuiper belt analogs around these stars. 

Most if not all, published {\it Spitzer} debris disk surveys have focused on either IRS spectra (14-34 $\micron$) or MIPS 24 and 70 $\micron$ photometry with several surveys reporting 160 $\micron$ data for young stars (Rebull et al. 2008; Cieza et al. 2008). While the 24 and 70 $\micron$ results suggest that 160 $\micron$ observations should often show excess emission, {\it Spitzer} is much less sensitive at this wavelength due to confusion from extra-galactic sources in its large beam and to an artifact in the data known as $``$the leak''.
This paper presents the results of our analysis of the 160 $\micron$ data taken as part of a {\it Spitzer} survey of nearby main sequence stars. Basic debris properties are derived from models of the spectral energy distribution of those stars with excesses at 160 $\micron$.  Section 2 briefly discusses the target sample considered in the study, $\S$ 3 reviews the reduction of the images, $\S$ 4 discusses the analysis used to remove the leak and determine the 160 $\micron$ photometry, $\S$ 5 summarizes the statistics of our findings and SED models of the data and $\S$ 6 reviews the implications of our results on the
overall properties of the disks around mature stars. 

\section{Sample} 

The primary sample of stars analyzed in this paper originates from the collection of mature FGK stars observed with MIPS at 24 and 70 $\micron$ (Beichman et al. 2005; Bryden et al. 2006, Program Identification (PID) \# 41). To aid with the analysis of the
leak, we also include 13 bright stars observed at 160 $\micron$ as part of an M dwarf program (G. Rieke, PID 52) or in a MIPS calibration program (B. Latter, PID 1709). There were no criteria on the spectral type of the stars used to model the leak since they are meant to be bright PSF templates and the 160 $\micron$ flux originates from the Rayleigh Jeans portion of their blackbody spectra. Tables~\ref{sample} and ~\ref{leakstars} list  the science targets and calibration stars. 

\section{Observations and Data Reduction}

All Spitzer observations were made between December 2003 and March 2006 with integration times of 3 and 10 seconds. The data reduction is based on the DAT software developed by the MIPS instrument team (Gordon et al. 2005) and is similar to that performed on the 24 and 70 $\micron$ data (Beichman et al. 2006; Trilling et al. 2007; Bryden et al. 2006).  We used images processed beyond the standard DAT software and then mosaicked from individual frames with half pixel subsampling (Stansberry et al. 2007).  The resulting pixel scale is 8$\arcsec$ pixel$^{-1}$. The calibration factor of  0.968 $\mu$Jy arcsec$^{-2}$ DN$^{-1}$ s$^{-1}$ and the appropriate aperture corrections for the 160 $\micron$ MIPS images were taken from Stansberry et al. (2007).

\section{Analysis}

\begin{figure}[htbp]
\epsscale{1.0}
\plotone{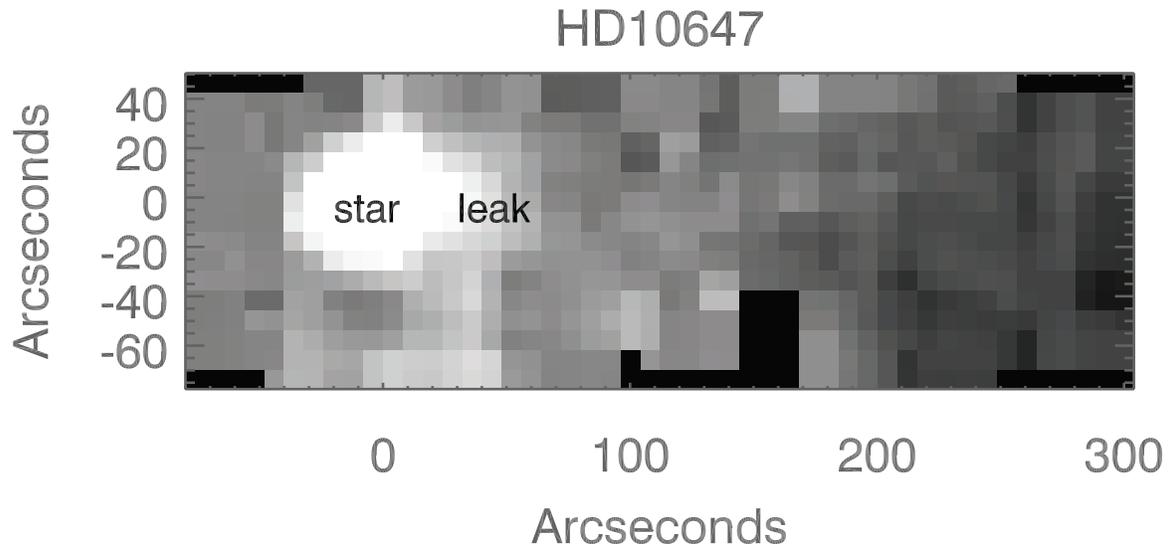}
\caption{Spitzer image of HD 10647 at 160 $\micron$ showing the position of the star relative to the leak. \label{hd10647}}
\end{figure}

The 160 $\micron$ data suffers from an imaging artifact (a.k.a. the $``$leak'') that is caused
by the reflection within the instrument of near-infrared light onto the MIPS detector (Stansberry et al. 2007, see Figure~\ref{hd10647}). The leak becomes an issue when looking for excesses around bright stars since the near-IR photosphere  is bright compared to the weak 160 $\micron$ signal.  The leak lies $\sim$4 pixels to the $+X$ side of the star's position on the detector and is close enough to hinder accurate 160 $\micron$ flux determinations with aperture photometry.

\begin{figure}
\epsscale{1.0}
\plotone{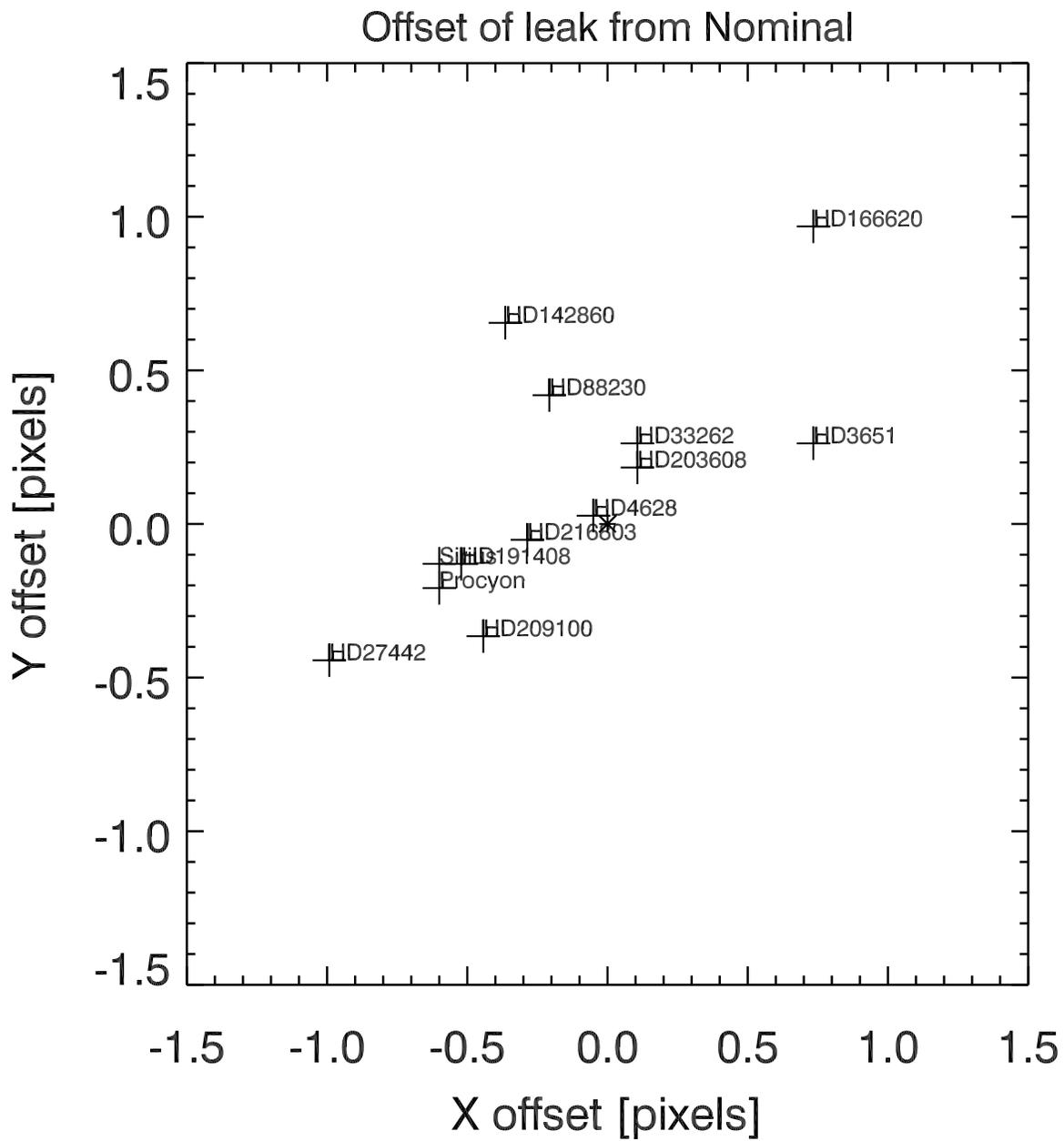}
\figcaption{ Plot of the variation in the offset in pixels between the leak centroid and the position of the star as estimated from the star's 24 $\micron$
position.    \label{offset}}
\end{figure}

Since the leak is due  to the near-infrared light from the star, we can
 predict the level of the leak emission based on the known near-IR brightness of the star.   We utilized archived {\it Spitzer} observations of a sample of stars which  have no 160 $\micron$ or shorter wavelength excess and which are bright in the near-infrared (see Table~\ref{leakstars}). These stars vary in brightness from $K_s=-1.39$ (Sirius) 
to $K_s=4.23$ (HD 166620).  We assume that the dominant source of flux in the 160 $\micron$ image near the position of the star is the leak. The offset of the 160 $\micron$ leak
from the position of the star, determined within a fraction  of an arcsecond  from the contemporaneous  24 $\micron$ MIPS image, has a position and dispersion of 4.5$\pm$1.5 pixels in the $X$ direction and -0.1$\pm$1.0 pixels in the $Y$ direction (see Figure~\ref{offset}). The spread in the leak offset is presumably due to differences in the observing parameters of the targets which have been collected from a number of different AORs.  

\begin{figure}
\epsscale{0.7}
\plotone{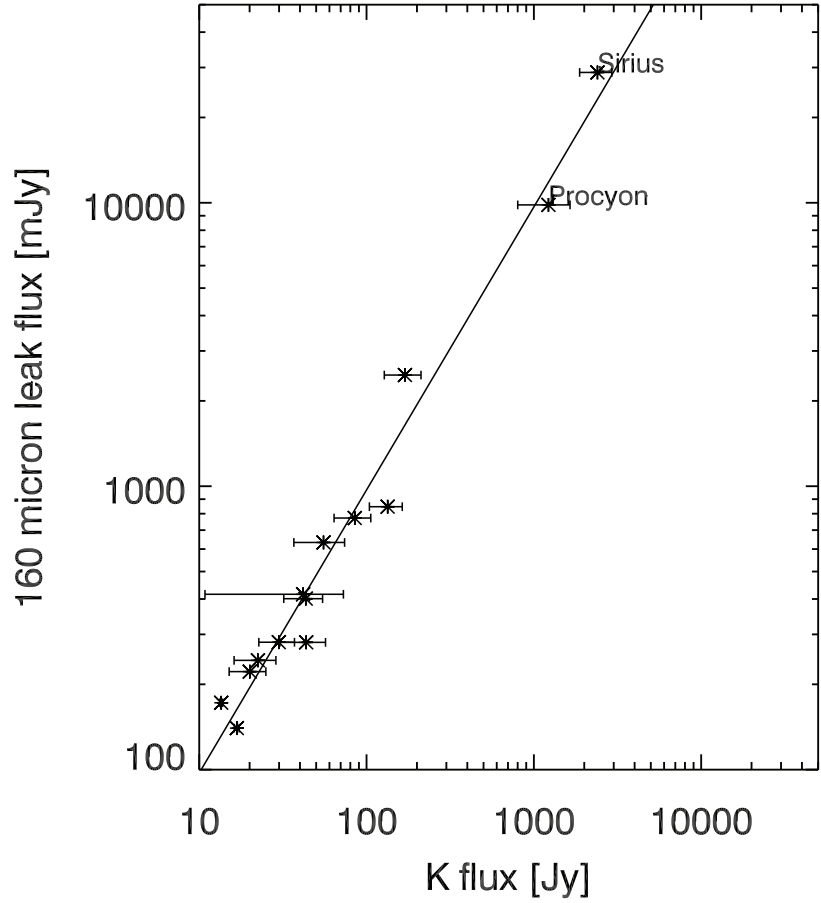}
\figcaption{ Plot of the K band flux from 2MASS versus the flux of the leak estimated from a sample of stars with 160 $\micron$ flux below
Spitzer sensitivities.  The plot shows a clear correlation between near-infrared flux and peak flux allowing for the potential of removing
the leak flux using this correlation as a guide. The best fitting line which is fit through all data points is $F_{leak [mJy]}=(10.5\pm1.0)F_{2.2 \micron [Jy]}+26\pm15$.
\label{kvsleak}}
\end{figure}

To estimate the flux of the leak we used aperture photometry with an aperture radius of 3 pixels (24$\arcsec$) and a sky annulus of 3-5 pixels (24-30$\arcsec$). Figure~\ref{kvsleak} plots the 160 $\micron$ leakage versus the 2MASS K$_S$ flux for the leak-only stars in our sample.  There is a clear dependence between the 160 $\micron$ leakage and K band flux with a correlation coefficient of 0.98.  The best fitting line through the  data  is 
 $F_{leak [mJy]}=(10.5\pm1.0)F_{2.2 \micron [Jy]}+26\pm15$. There is no correlation between the leak flux or offset and the J-K color of the star. 

\begin{figure}
\epsscale{1.0}
\plottwo{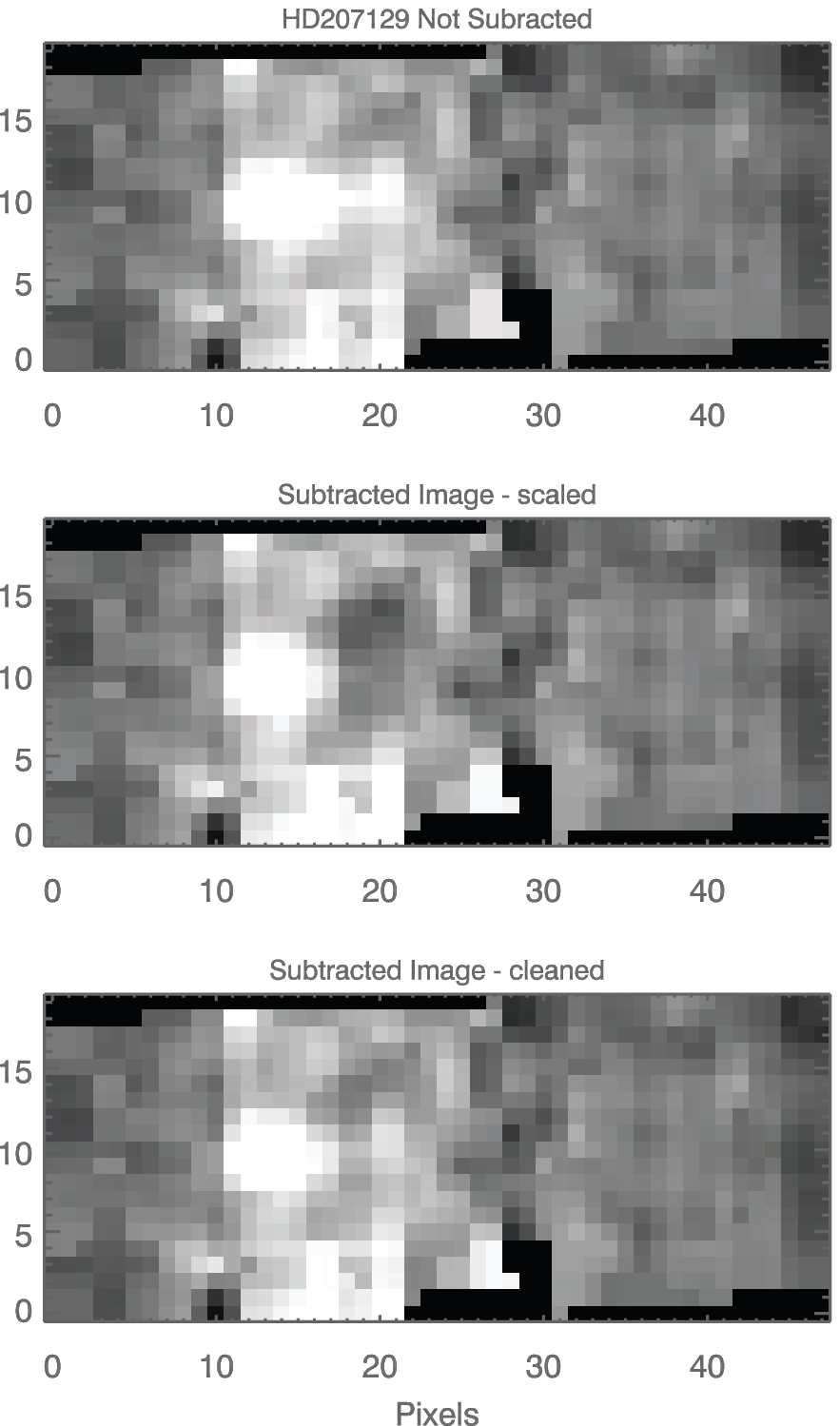}{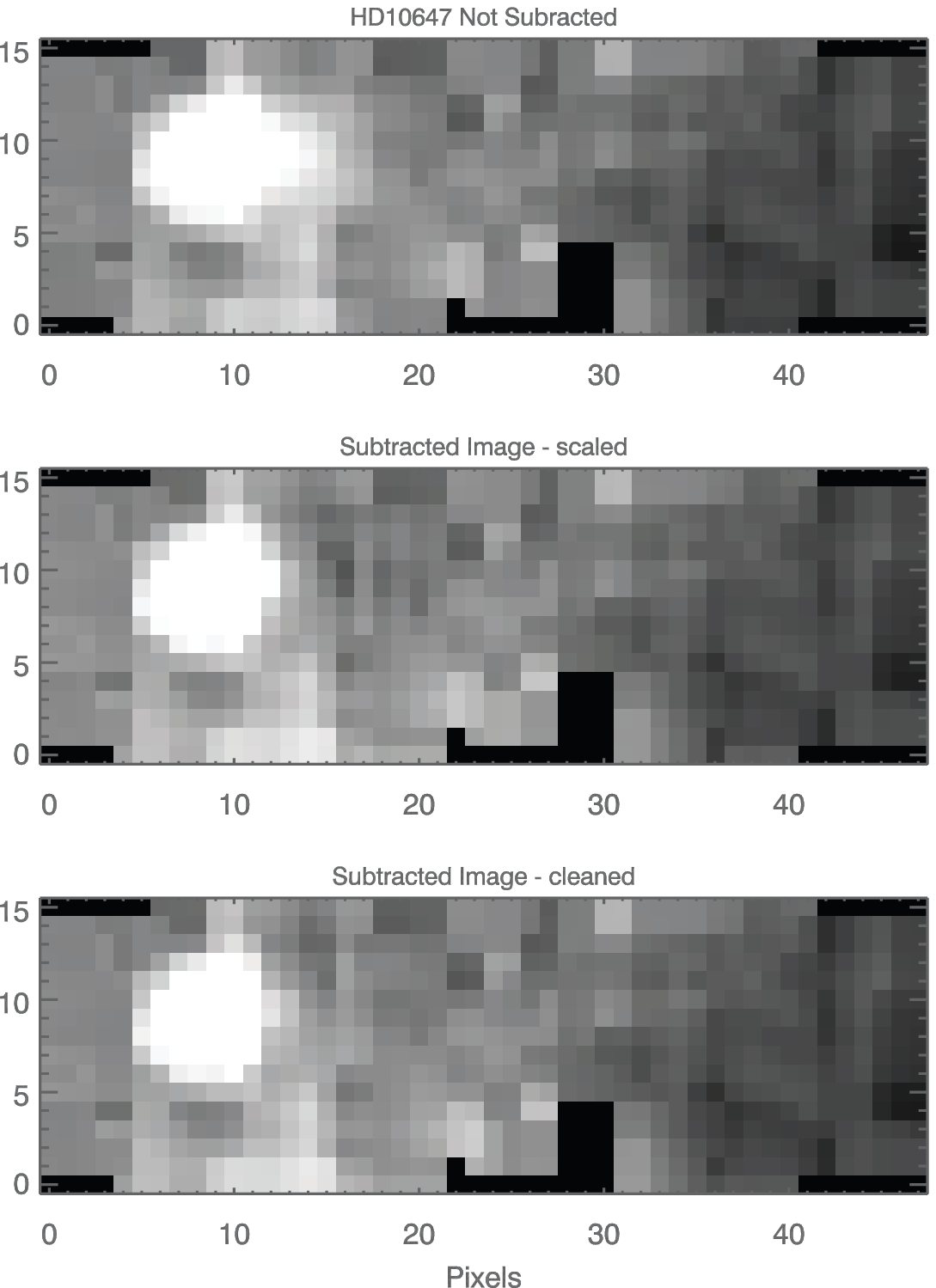}
\figcaption{ MIPS 160 $\micron$ images of HD 10647 and HD 207129 both before (top) and after the leak was removed using the 
scaled leak (middle) and iterative (bottom) cleaning processes.  The plate scale of the images
is 8'' pixel$^{-1}$.
 \label{cleaned}}
\end{figure}

To remove the leak from our target sample,  we create  a ``leak'' Point Spread Function (PSF) by scaling and averaging the four brightest leak-only images (Sirius, Procyon, HD 80007, and HD 27442). The derived ``leak'' PSF is shifted to the appropriate position, scaled to  the predicted  flux  and subtracted from the image. Figure~\ref{cleaned} (middle pane) shows the result of this subtraction for HD 10647 and HD 207129.  In some cases the leak was not fully removed using the scaled  PSF. This could be due to large errors in the K$_S$ band flux due to saturation in the 2MASS images as suggested by the K$_S$ error bars  shown in Figure~\ref{kvsleak}. 

\begin{figure}
\epsscale{0.7}
\plotone{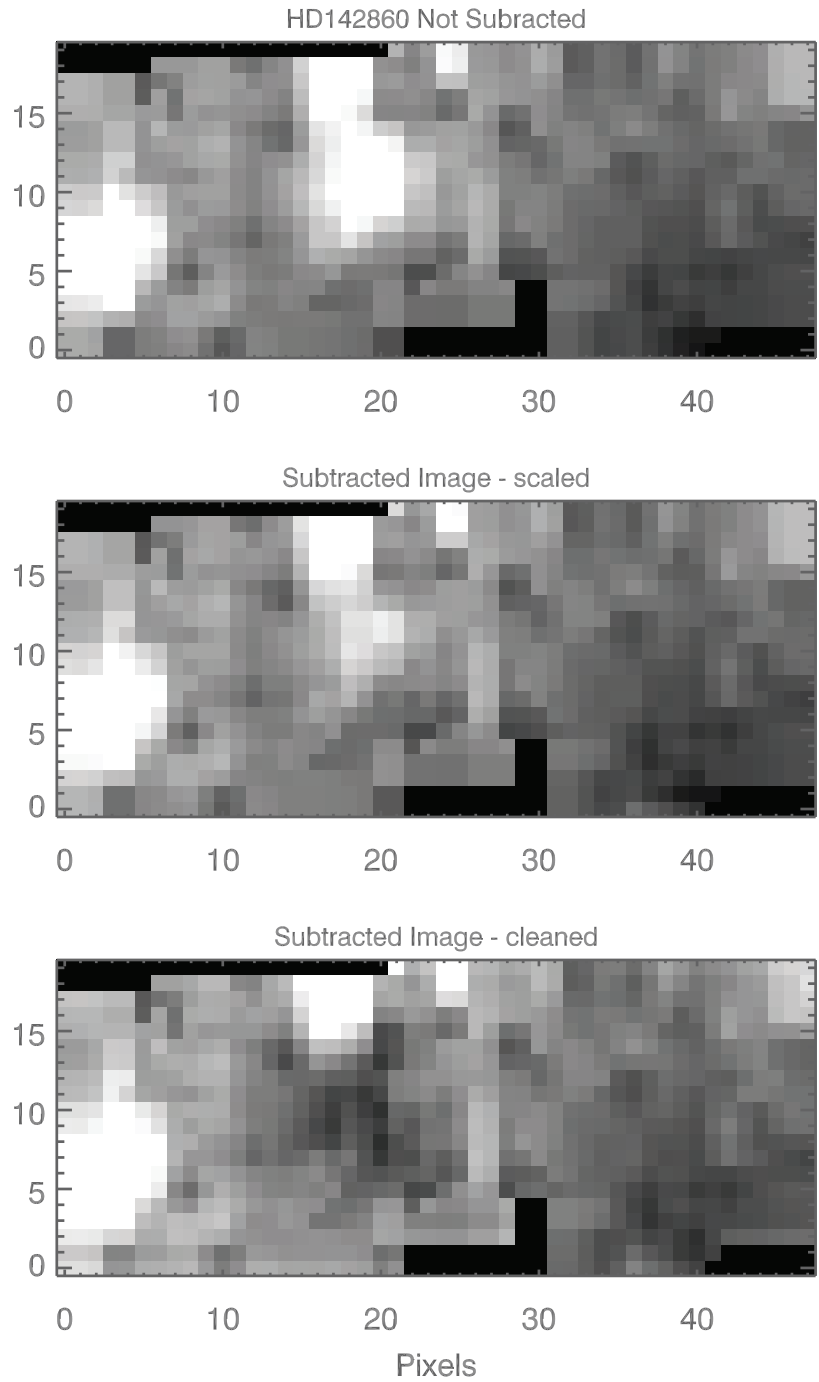}
\figcaption{ MIPS 160 $\micron$ images of HD 142860 before (top) and after the leak was removed using the 
scaled leak (middle) and iterative (bottom) cleaning processes. In this case, there is no detection at 160 $\micron$ but
its clear that the leak has been fully removed from the image using the iterative method. The plate scale of the images
is 8'' pixel$^{-1}$.
 \label{nondet}}
\end{figure}

As an alternative method to remove the leak, we employ a cleaning method which subtracts a scaled version of the leak iteratively in order to reduce  the signal level in the leak region to match that in the surrounding background.   First, a circular region with a radius of four pixels at the position of the leak is declared the cleaning area. During the cleaning
process, the highest point within the cleaning region is identified and the leak PSF is scaled to 10\% of the value of the pixel at that position. This scaled leak
PSF is subtracted at this position and then the process is repeated until the RMS noise 
within the cleaning area is at the same level as that in an annulus around the position of the star. Figure~\ref{cleaned} (bottom pane) shows the results of the iterative method for HD 10647 and HD 207129. Figure~\ref{nondet} shows the results of the leak removal process for HD 142860 which
does not have significant excess emission at 160 $\micron$. These cleaned images show that the leakage  is getting reduced to the level of the background using the iterative method. We use the results from both the scaled leak and the iterative cleaning methods in determining the uncertainty in the 160 $\micron$ flux.

\begin{figure}
\epsscale{1.0}
\plotone{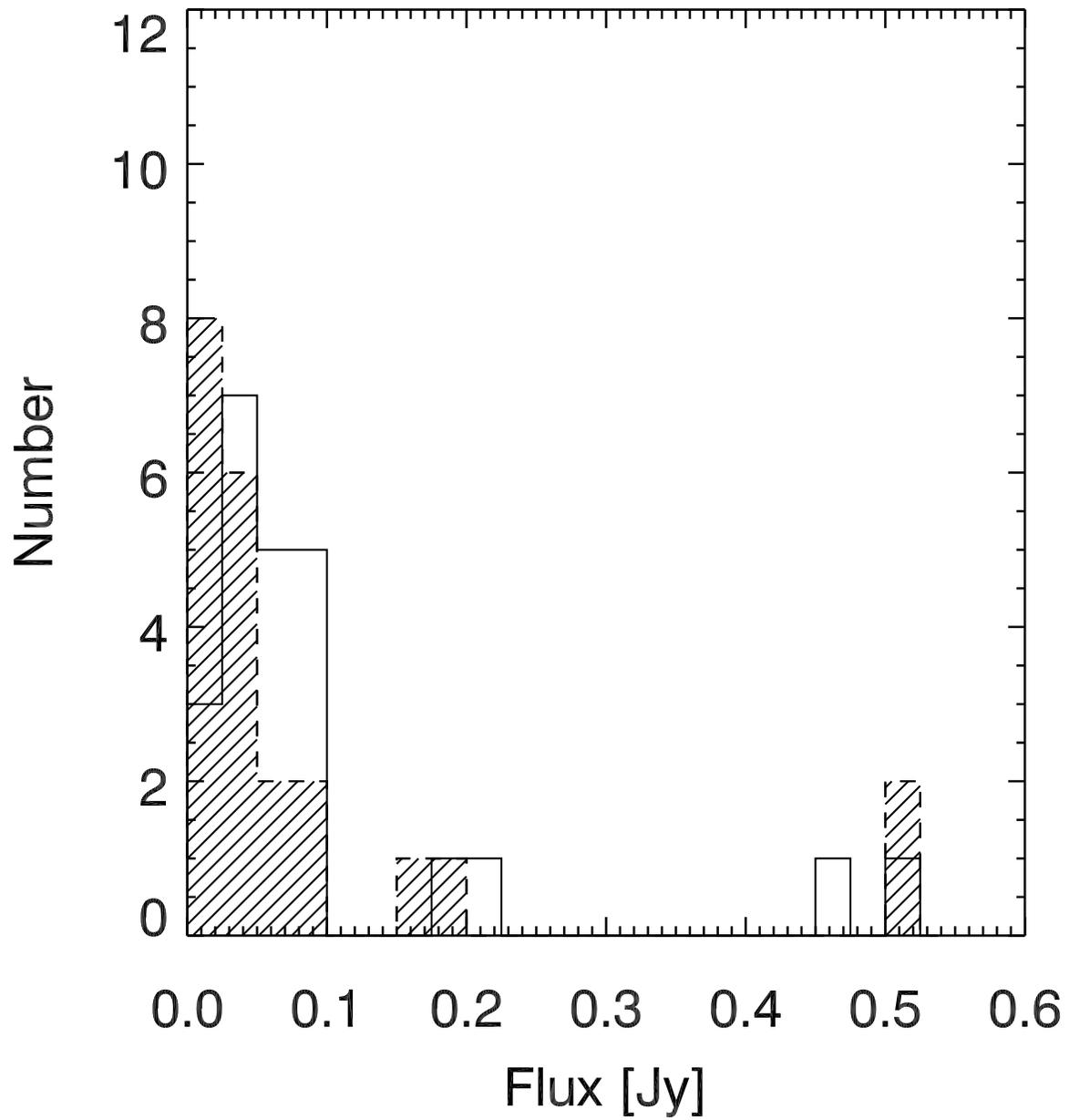}
\figcaption{Histogram of the background flux density at the positions of the 160 $\micron$ stars before (empty) and after (diagonal) leak removal. 
Removing the leak results in an 40\% improvement in the flux sensitivity of the data based on the average values of these two distributions.   \label{fluxhist}}
\end{figure}

We performed aperture photometry on the leak-removed and raw images using an aperture of 4 pixels (32$\arcsec$) in radius and a sky annulus of 7-8 pixels (56-64$\arcsec$).  
 Figure~\ref{fluxhist} shows a histogram of the flux densities measured within the aperture before (empty) and after (diagonal) removal of the leak for all stars in the 160 $\micron$ sample. 
Removing the leak results in an 40\% improvement in the flux sensitivity of the data 
based on the average values of these two distributions. For those stars with no significant 160 $\micron$ excess, we estimate an upper limit to the 160 $\micron$ photometry based on the flux that would be necessary for an excess to be detected given our detection criteria and background noise estimate for each image. Table~\ref{results}  lists
the upper limits for those stars with 70 $\micron$ excesses and no 160 $\micron$ excesses. 
Most of these upper limits provide a more stringent constraint on the range of temperatures expected for the cold dust population that would be estimated from solely the 70 $\micron$ detection.

\section{Results}

For this study, we use the criterion that a star has a 160 $\micron$ excess if the flux at the position of the star after leak removal and subtraction of the stellar photosphere is 3 $\sigma$ above the background noise in the image. To determine the level of background noise, we first correct for the flux gradient apparent over the length of the MIPS 160 array. 
The gradient is removed by fitting a line to pixels at both ends of the array and using the slope to create a gradient which is subtracted from the image. The background noise is then estimated from the standard deviation of the counts in eight apertures placed 14 pixels (122$\arcsec$)  to the right ($+X$) of the star. Using these criteria, only three stars have significant 160 $\micron$ emission after subtraction of the leak: HD 10647, HD 207129, and HD 115617. Table~\ref{excesses} lists the 160 $\micron$ flux densities estimated with aperture photometry from the non-leak subtracted image, the image with the K-band scaled leak subtracted and the image with the leak removed iteratively. We also list the background noise  flux density and the flux density of the stellar photosphere at 160 $\micron$.
Unlike the 24 and 70 $\micron$ studies, all detections at 160 $\micron$ are dominated by the excess emission above the stellar photosphere.

\begin{figure}
\epsscale{1.0}
\plotone{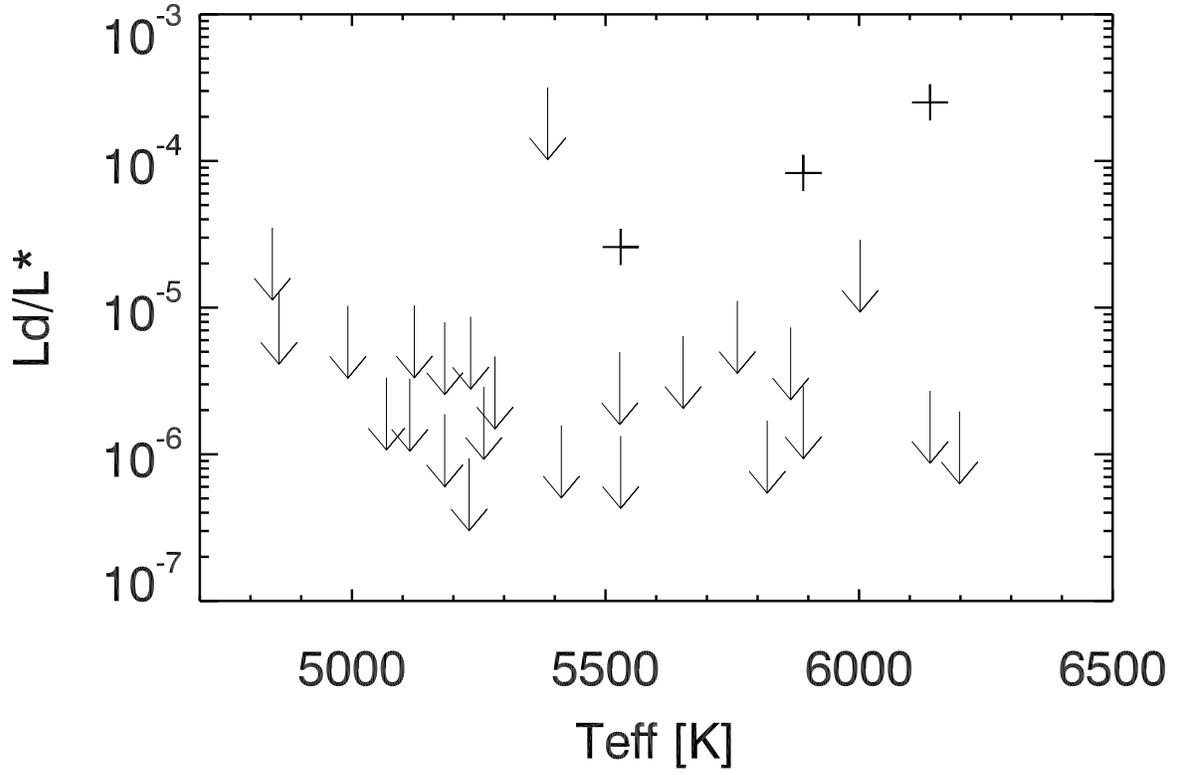}
\figcaption{ Plot of the limits on L$_d$/L$_*$ for all stars in the 160 $\micron$ sample as a function of stellar effective temperature. The crosses, +, are those stars with
detected 160 $\micron$ flux. Those stars with non-detections have upper limits in their L$_d$/L$_*$ values provided by Equation 2.  \label{sens_hist}}
\end{figure}

Figure~\ref{sens_hist} plots the minimum detectable disk luminosity, L$_d$/L$_*$, estimated from the 160 $\micron$ excess as a function of stellar effective temperature, T$_*$. 
This assumes a single disk temperature which is determined by setting
the blackbody emission peak at 160 $\micron$ corresponding to a dust temperature of 23 K.  The minimum disk luminosity is then the ratio of the total flux determined from this blackbody
to the stellar photospheric flux (see Equation 1, where F$_{160}$ is the 160 $\micron$ flux density and F$_{160,*}$ is the flux density of the stellar photosphere at 160 $\micron$). Most of these luminosity limits (see Table~\ref{dustprop}) are an improvement over those provided by 70 $\micron$ limits from previous studies (Bryden et al. 2006). 

\begin{equation}
\frac{L_{dust min}}{L_{*}} = 8\times10^{-7}  \left(\frac{5600 K}{T_{*}}\right)^3   \frac{F_{160}}{F_{160,*}}
\end{equation}

\noindent Using these  L$_d$/L$_*$ values, we can provide an estimate of the dust mass assuming 1 $\micron$ radius grains with a density of 3.3 g cm$^{-3}$ at a distance of 100 AU 
(Jura et al. 1995, see Table~\ref{dustprop}).

\begin{figure}
\epsscale{1.}
\plottwo{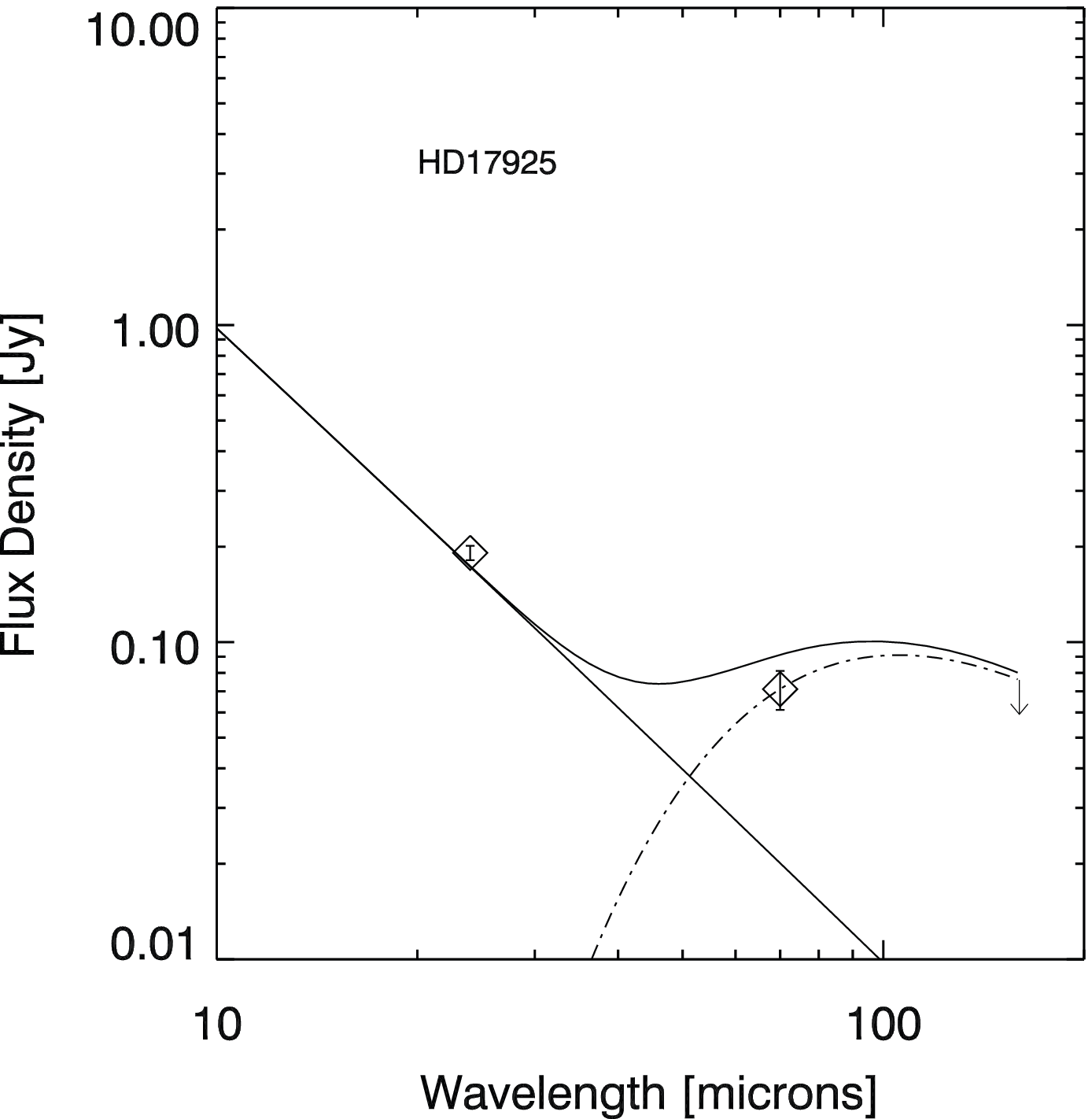}{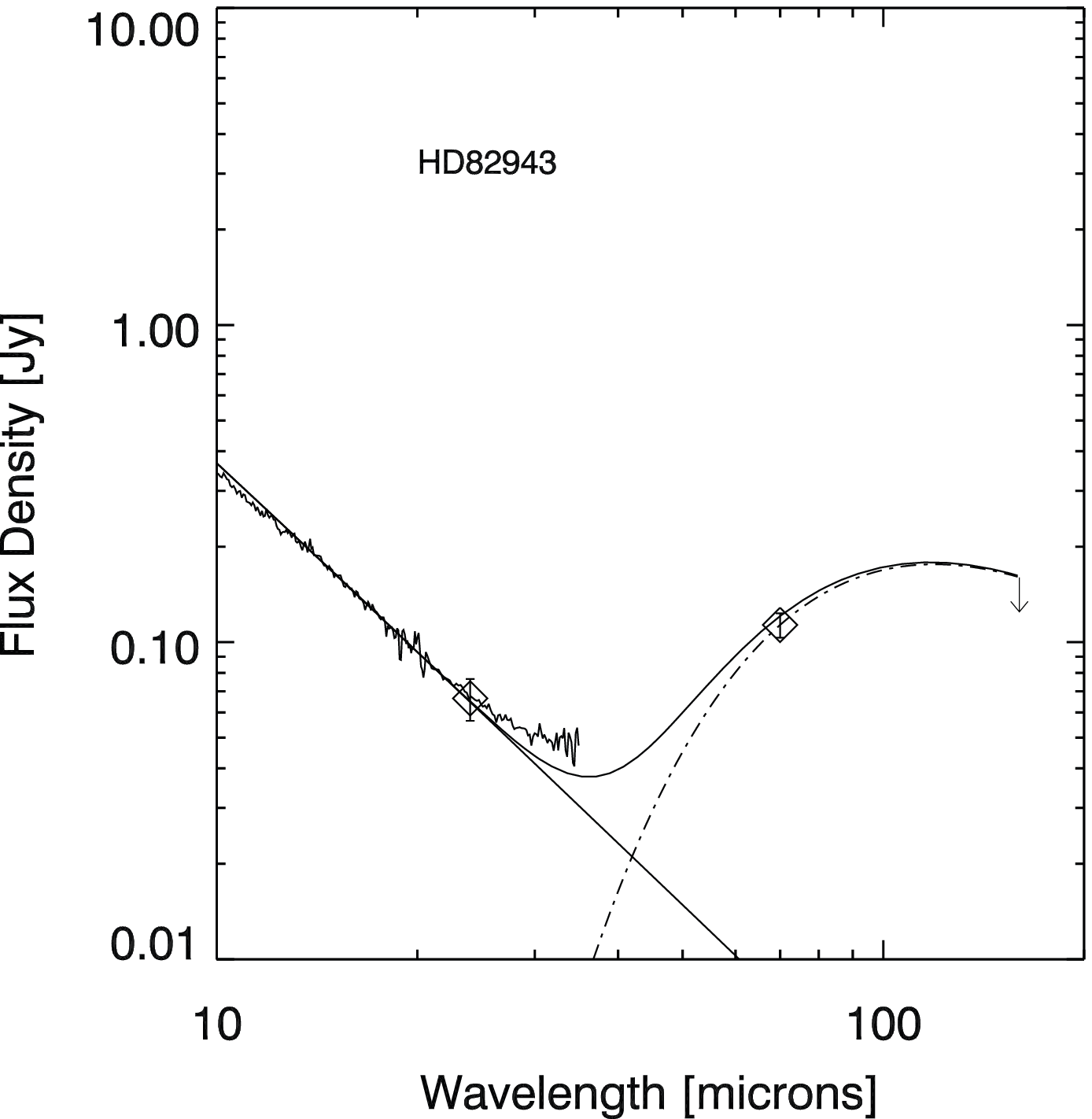}
\figcaption{ SEDs of HD 17925 and HD 82942 which do not have significant 160 $\micron$ excesses. The upper limit in 160 $\micron$ flux density can
still be used to constrain the properties of the disk. \label{seds2}}
\end{figure}

For those stars with 70 $\micron$ excesses and 160 $\micron$ upper limits, we can estimate a lower limit for the temperature of the dust component responsible for the excess at these two wavelengths by fitting a blackbody curve to these two data points (see Table~\ref{dustprop}). Unfortunately, these temperature limits are more indicative of the variations in the background levels of the 160 $\micron$ images than the properties of the dust. As an example, Figure~\ref{seds2} shows spectral energy distributions of HD 17925 and HD 82943 with a blackbody fit to their 70 $\micron$ excess and 160 $\micron$ upper limit. Equation 2 shows that we can also estimate a limit on L$_d$/L$_*$ (also in Table~\ref{dustproplim}) in the same manner as for the 160 $\micron$ detections but with the 3 $\sigma$ flux limits. 

\begin{equation}
\frac{L_{dust}}{L_{*}} < 8\times10^{-7}  \left( \frac{5600 K}{T_{*}}\right) ^3   \frac{3\, N_{160}}{F_{160,*}}
\end{equation}

\noindent where T$_{*}$ and F$_{160,*}$ are the stellar effective temperature and flux, respectively, and N$_{160}$ is the one $\sigma$ flux density upper limit. Finally, we can estimate the smallest amount of dust that could be present to have no 160 $\micron$
detection. Using the  L$_d$/L$_*$ lower limits, we can provide a lower limit to the dust mass using the same assumptions for those stars with 160 $\micron$ excesses.  Table~\ref{dustproplim} lists these dust masses as fractions of an Earth mass. 

\subsection{Spectral Energy Distributions}

Here we employ simple SED models to fit the 24, 70 and 160 $\micron$ MIPS photometry and IRS spectra from previous studies of the three stars with 160 $\micron$ excesses. 
The 24 and 70 $\micron$ photometry comes from Trilling et al. (2008) and the IRS data come from either Beichman et al. (2006) or Lawler et al. (2009). All of the stars with 160 $\micron$ detections have disks which are resolved at 70 $\micron$ (HD 10647, 106 AU (radius); HD 207129, 144 AU; and HD 115617, 97 AU; Bryden et al. 2009 in prep). The 70 $\micron$ sizes for HD 10647 and HD 207129 both agree with the Hubble Space Telescope (HST) measurements (Stapelfeldt et al. 2007) which detect a relatively narrow toroidal structure at 90-140 AU using the Advanced Camera for Surveys (ACS). 

\subsection{Silicates or Black Body Dust Models}

To model the debris disk we place the dust in an optically thin annulus around the star and utilize the formalism of Su et al. (2005). 
The dust temperature varies with the distance, $r$, from the star as determined from radiative equilibrium. 
The free parameters in the SED model fits are the inner and outer radius of the dust annulus, $r_i$ and $r_o$, and the dust density at the inner radius which we
later convert to L$_d$/L$_*$ and then the total dust mass in the system. Also, we make assumptions of the dust composition which affects the shape of the SED 
We choose to fit the {\it Spitzer} data with models incorporating either 
blackbody grains, small silicate dust grains ($a=0.25$ $\micron$) or large silicate dust grains ($a=10$ $\micron$), with 
emission coefficients taken from Laor \& Draine (1993). The smallest silicate grains used in our models are bigger than the
blowout radius of the grain which depends on the luminosity of the star.  Since 10 $\micron$ dust grains are primarily affected
by Poynting-Robinson drag, we model this disk component with a constant surface density. The 0.25 $\micron$ grains, on the other hand, are primarily removed from the disk by radiation pressure
resulting in a surface density with a r$^{-1}$ fall off (Su et al. 2005). If we include the additional constraint that the 70 $\micron$ MIPS photometry must be modeled with a dust component 
with an inner radius corresponding to the resolved 70 $\micron$ radius, then 
the Spitzer data is best fitted with two separate dust components with different dust species.

\begin{figure}
\epsscale{1.0}
\plotone{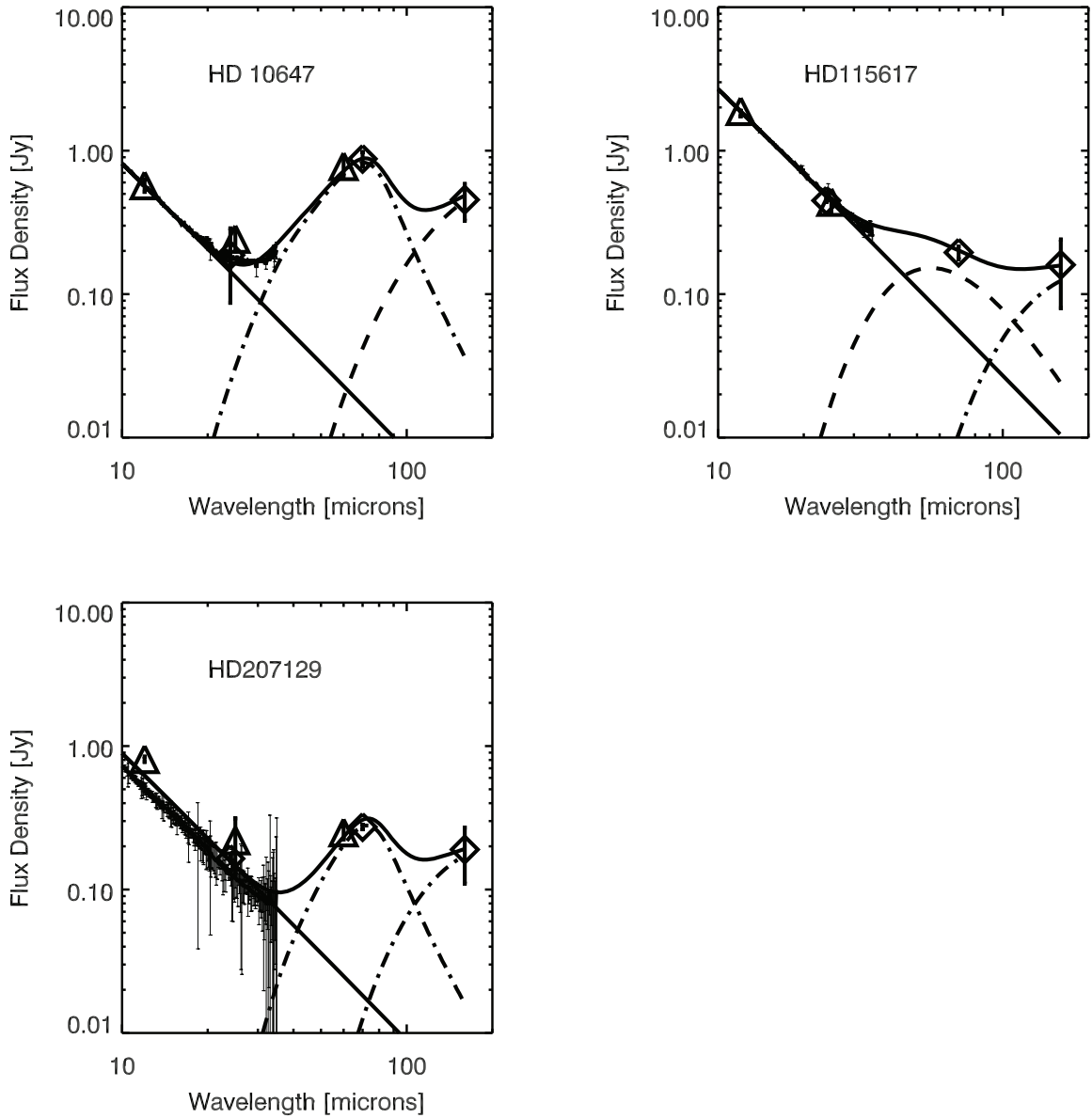}
\figcaption{SEDs of those stars with 160 $\micron$ excesses. In all cases, these stars also have excesses at 70 microns
allowing for a fit to the temperature of the dust. The dust is placed in an annulus around the star with a number density fall off dependent on the dust removal mechanism. 
The additional constraint of the extent of the dust at 70 $\micron$ and the IRS spectra, results in the best fit (solid line) being achieved
with two separate dust annuli (dotted and dashed lines) of varying composition. The dust temperatures (20-70 K) were estimated assuming radiative equilibrium and depended on the 
dust composition, stellar luminosity and distance of the dust from the star). The properties of the dust
derived from these SED model fits are listed in Table~\ref{dustprop}. 
The triangles are IRAS (25, 60 $\micron$) fluxes and the squares are the MIPS data.  \label{seds}}
\end{figure}

The properties of the dust derived from the best fits to the {\it Spitzer} data (the inner and outer radii of the dust annulus, L$_d$/L$_*$,  and the dust composition) are listed in Table~\ref{dustprop} along with 
derived dust properties such as dust temperature and dust mass. 
Figure~\ref{seds} shows the best fitting SED models for all three stars with 160 $\micron$ excesses.
The errors on the variables used to fit the SED, the inner and outer dust radius and inner dust density of each dust component, 
were estimated from the change in their value which produced a variation of one in
the $\chi^2$ (Press et al. 2002). 

\subsection{Icy Dust Models}

\begin{figure}
\epsscale{1.0}
\plotone{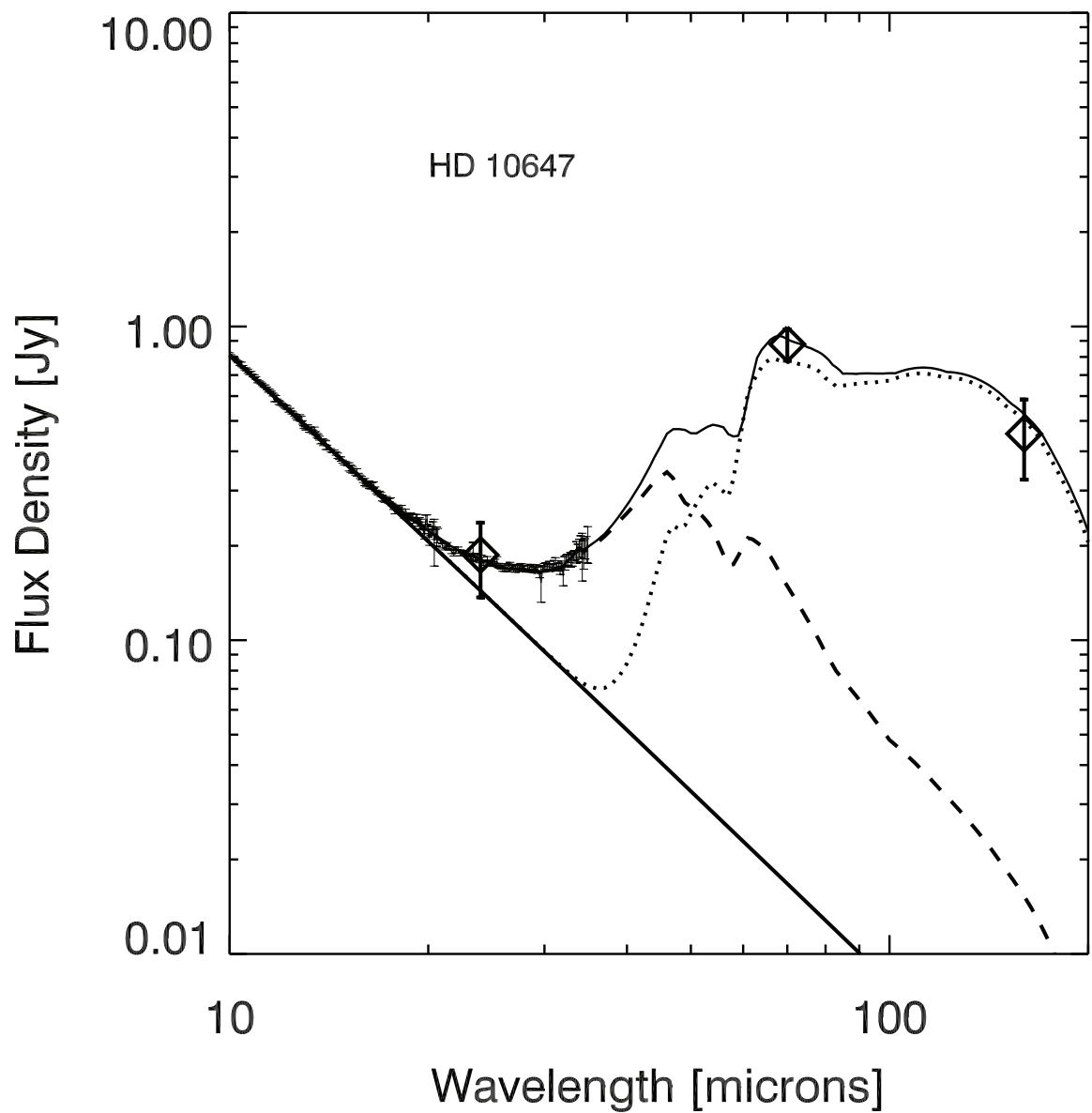}
\figcaption{Plot of an additional model of the Spitzer data for HD 10647. This model uses 30 K water ice particles to model the 70 and 160 $\micron$ emission component (dotted)
and $\sim$70 K silicates to model the IRS data (dashed). 
The model has been normalized to the MIPS 24 $\micron$ flux density. 
 \label{lisse2}}
\end{figure}

A more thorough investigation of the SED of HD 10647, based on the primitive solar system dust models of Lisse et al. (2007, 2008) reveals that,
after modeling the IRS 5-35 $\micron$ warm dust signature, the 70 and 160 $\micron$ flux densities can be modeled by a component of $\sim$30 K water ice particles
at 100 AU from the star. An enhanced peak in the 70 $\micron$ passband occurs due to emission from a strong water ice vibrational feature at 65 $\micron$ (see Figure~\ref{lisse2}). While additional contributions to the dust emission in the IRS
portion of the spectrum could be due to silicates, our data set cannot place statistically significant constraints on the abundance of such materials. These SED models do, however, showcase the
improvement in our ability to learn about the dust with continuous spectral coverage and direct spatial information. A distance of $\sim$100 AU from the primary is consistent with the 90 - 140 AU location of dust detected in the optical with HST/ACS (Stapelfeldt et al. 2007).

In addition to the Spitzer observations presented here, thermal emission from the HD 10647 circumstellar dust has been studied most recently at 870 $\micron$ using the sub-mm LABOCA array on the 12-m APEX telescope in Chile (Liseau et al. 2008). In order to fit the 870 $\micron$ flux point of 39$\pm$8 mJy for this star, the dominant (by surface area) water ice particle sizes must be on the order of 1 mm. However, as stated in Liseau et al. (2008), the morphology of this cold component is very broad, on the order of 600 AU FWHM. It is most likely that the material emitting the bulk of the 870 $\micron$ emission is at 17 K, and is unassociated with the material constrained to the 100 AU torus. We are unable to fit the 870 $\micron$ data point with the water ice SED component at 30 K which is consistent with the hypothesis of an  additional 17 K dust component at a distance of $\sim$ 300 AU. 

\section{Discussion}

For this sample we only consider those stars in the original FGK survey (PID41, Beichman et al. 2005; Trilling et al. 2007) with 160 $\micron$ data. While all stars with 160 $\micron$ excesses also have 70 $\micron$ excesses,  not all stars with 70 $\micron$ excesses have 160 $\micron$ excesses.
Out of 35 stars in the FGK sample, two have 160 $\micron$ excesses. The other one, HD 10647, was included in a separate survey with different goals. 
Twelve of the 160 $\micron$ non-detections have 70 $\micron$ excesses. One (HD 10647) of the five stars with known RV planets has a 160 $\micron$
excess (20\%), compared to 2 out of 31 (6\%) of the stars with no known planets. This sample is smaller than the 24/70 $\micron$ samples so these percentages suffer from small number statistics.  
Given the 160 $\micron$ sensitivity limits due to confusion noise and the leak, we are generally not sensitive to 
dust much colder than 25 K and 0.1$\times$10$^{-4}$ M$_{\earth}$ (see Table~\ref{dustprop}).

A collection of icy dust with a mass of $\sim$4$\times$10$^{-4}$ M$_{\earth}$ at 30 K around HD 10647 is very reasonable with respect to known structures in the solar system. Dust created by interactions in the Saturnian and Uranian systems (such as formed the system's rings) will equilibrate at 60-90 K. Dust created by collisional fragmentation of bodies in the Kuiper Belt (Davis \& Farinella 1997; Stern 1995) produces originally toroidal morphologies at local temperatures of 35-45 K.  The location of the dust is far removed from the orbit of the reported planet HD 10647b with semi-major axis of 2.1 AU (Mayor et al. 2003) and is not likely to be dynamically coupled to  it.

\section{Conclusions}

We have analyzed the 160 $\micron$ subset of a survey of nearby main sequence stars with the {\it Spitzer}  IRS and MIPS instruments. After careful analysis of the 160 $\micron$ images which suffer from an anomalous flux contaminant, we have found three stars (HD 10647, HD 207129 and HD 115617) with significant excesses in this passband.  
All of these stars also have excesses at 70 $\micron$ and
the addition of the 160 $\micron$ photometry allows for a constrained estimate of the temperature of the dust as well as its composition. The lack of 160 $\micron$ emission for the remaining stars also sets an improved limit to the amount of cold dust for those sources with 70 $\micron$ excesses. In the end, the sensitivity floor of the 160 $\micron$ images is set by the background noise due to extragalactic sources. 

The Hershel Space Telescope, scheduled to launch in 2009, will have about four times better resolution than {\it Spitzer} and will be $\sim$10 times more sensitive at
comparable wavelengths suggesting we might be able to resolve those disks detected at 160 $\micron$. Having the intrinsic size of the debris disks as well as robust fluxes at
multiple wavelengths would further constrain the SED models. Those stars with 70 $\micron$ excesses and no 160 $\micron$ detections are ideal targets as well since there is dust emission at longer wavelengths not detectable with {\it Spitzer}. 

\begin{acknowledgements}
The research described in this 
publication was carried out at the Jet Propulsion Laboratory,
California Institute of Technology, under a contract with the National 
Aeronautics and Space Administration. This publication makes use of data products 
from the Two Micron All Sky Survey, which is a joint project of the University of 
Massachusetts and the Infrared Processing and Analysis Center/California Institute 
of Technology, funded by the National Aeronautics and Space Administration and 
the National Science Foundation.
\end{acknowledgements}

\begin{deluxetable}{lccccc}											
\tablecaption{Stars in the 160 $\micron$ Sample \label{sample}}	
\footnotesize										
\tablehead{											
\colhead{Star}	&	\colhead{SpTy}	&	\colhead{K$_S$}	&	\colhead{Distance [pc]}	&	\colhead{PID$^a$}	 
}											
\startdata											
HD 166	       &	 K0Ve	&	6.13	&	13.70	&	41	\\
HD 10476	&	 K1V	         &	5.20	&	7.47	         &	41	\\
HD 10647	$^b$&	 F9V   	&	5.52	&	17.35	&	716	\\
HD 13445$^b$	&	 K1V          &	6.17	&	10.91	&	41	\\
HD 17925	&	 K1V  	&	6.00	&	10.38	&		41	\\
HD 33262	&	 F7V   	&	4.72	&	11.65	&		41	\\
HD 33636	&	 G0	         &	7.06	&	28.69	&	         41	\\
HD 37394	&	 K1V  	&	6.23	&	12.24	&		41	\\
HD 52265$^b$	&	 G0III-IV	&	6.30 	&	28.07	&	        41	\\
HD 72905	&	 G1.5Vb	&	5.65	&	14.27	&		41	\\
HD 76151	&	 G2V	         &	6.00	&	17.09	&		41	\\
HD 82943$^b$	&	 G0      	&	6.54	&	27.46	&		41	\\
HD 114783$^b$	&	 K0	&	7.57	&	20.43	&	41	\\
HD 115617	&	 G5V	&	4.74	&	8.53	&		41	\\
HD 117176	&	 G5V	&	5.00	&	18.11	&		41	\\
HD 118972	&	 K2V	&	6.93	&	15.61	&		41	\\
HD 128311	&	 K0V	&	7.51	&	16.57	&		41	\\
HD 145675	&	 K0V	&	6.67	&	18.15	&        	41	\\
HD 149661	&	 K2V	&	5.76	&	9.78	&		41	\\
HD 177830$^b$	&	 K0	&	7.18	&	59.03	&     41	\\
HD 185144	&	 K0V	&	4.70	&	5.77	&		41	\\
HD 190007	&	 K4V	&	7.48	&	13.11	&	41	\\
HD 207129	&	 G0V	&	5.58	&	15.64	&	41	\\
HD 219134	&	 K3V	&	5.56	&	6.53	&	41	\\
HD 220182	&	 K1V	&	7.36	&	21.92	&	41	\\
\enddata		
\tablenotetext{a}{PID = program identification number}	
\tablenotetext{b}{Stars with known planet(s).}								
\end{deluxetable}	

\begin{deluxetable}{lccccc}											
\tablecaption{Stars Used for the Leak PSF \label{leakstars}}	
\footnotesize										
\tablehead{											
\colhead{Star}	&	\colhead{SpTy}	&	\colhead{V}	& \colhead{K$_S$} &	\colhead{Distance [pc]}	&	\colhead{PID$^a$}	 
}											
\startdata											
HD 3651	&	 K0V	         &	5.80	& 4.00$\pm$0.04 &	11.11	&	41	\\
HD 4628	&	 K2V  	&	5.75	& 3.68$\pm$0.27 &	7.46	         &	41	\\
HD 27442	&	 K2IVa	&	4.44	& 1.75$\pm$0.22 &	18.23	&		41	\\
HD 80007	&	 A2IV	&	1.70	&1.49 $\pm$0.24 &	34.08	&	1709	\\
HD 88230	&	 K5V	         &	6.61	& 2.96$\pm$0.29 &	4.87	&	41	\\
HD 142860	&	 F6IV	&	3.85	& 2.70$\pm$0.31 &	11.12	&        	41	\\
HD 166620	&	 K2V	&	6.37	& 4.23$\pm$0.02 &	11.10	&	41	\\
HD 191408	&	 K3V	&	5.31	& 3.01$\pm$0.60 &	6.05	&	41	\\
HD 203608	&	 F6V	&	4.22	& 2.97$\pm$0.25 &	9.22	&	41	\\
HD 209100	&	 K4.5V&	4.69	& 2.24$\pm$0.24 &	3.63	&	41	\\
HD 216803	&	 K4V	&	6.48	& 3.81$\pm$0.24 &	7.64	&	41	\\
Procyon	&	 F5IV-V	&	0.34	& -0.66$\pm$0.32 &	3.50	&     52	\\
Sirius	&	 A1V	         &	-1.47	& -1.39$\pm$0.21 &	2.64	&	52	\\
\enddata		
\tablenotetext{a}{PID = program identification number}									
\end{deluxetable}

\begin{footnotesize}
\begin{deluxetable}{lcccccc}		
\footnotesize							
\tablecaption{Flux Densities for Stars with 160 $\micron$ excesses \label{excesses}}									
\tablehead{									
\colhead{Star}	&	\colhead{Raw$^a$ }	&	\colhead{Excess - scaled leak method$^b$}	&	\colhead{Excess - iterative method$^b$} & \colhead{Noise$^c$} & \colhead{Significance$^d$} & \colhead{Stellar Photosphere$^e$}  \\
                            &     \colhead{mJy}          &    \colhead{mJy}                            &     \colhead{mJy}               &      \colhead{mJy}         &   \colhead{mJy} &    \colhead{mJy}                    
}									
\startdata									
HD 10647  	&	498	&  462    &  451     & 50	& 9.2   & 3.17  \\
HD 207129	&	185	& 152	& 158   & 20	& 7.9   & 3.45 \\
HD 115617	&	318	& 141	& 89     & 20	& 4.5     &10.56  \\
\enddata		
\tablenotetext{a}{The raw flux represents the combination of the flux from the leak, the stellar photosphere and excess thermal emission.}
\tablenotetext{b}{The "scaled leak method" and "iterative method" represent two different forms of analysis utilized to removed the leak flux prior to performing aperture photometry at the position of the star. The
difference in the two methods are described in Section 4. }
\tablenotetext{c}{The background noise is estimated from the standard deviation on the flux from eight apertures placed 35 pixels in the +x direction away from the star. }
\tablenotetext{d}{The significance given here is defined as the excess thermal flux (sky subtracted) from the iteratively cleaned image over the background noise.}	
\tablenotetext{e}{Value of the stellar photosphere at 160 $\micron$. This has been subtracted from all of the excess flux values.}						
\end{deluxetable}	
\end{footnotesize}
										
\begin{deluxetable}{lcc}											
\tablecaption{Stars With No 160 $\micron$ Excess \label{results}  }	
\tablehead{											
\colhead{Star}	 & \colhead{ Flux Density Limit [mJy] }	&	\colhead{70 $\micron$ excess?}
}											
\startdata											
HD 166          &	$<$30	&	yes	\\
HD 17925	&	$<$76	&	yes	\\
HD 33262	&	$<$150	&	yes	\\
HD 33636	&	$<$82	&	yes	\\
HD 37394	&	$<$76	&	yes	\\
HD 52265	&	$<$523	&	yes	\\
HD 72905	&	$<$55	&	yes	\\
HD 76151	&	$<$150	&	yes	\\
HD 82943	&	$<$160	&	yes	\\
HD 117176     &   $<$300	&	yes	\\
HD 118972     &	$<$110	&	yes	\\
HD 128311     &	$<$83	&	yes	\\
\hline											
HD 10476	  &	$<$90	&	no	\\
HD 13445	  &  $<$20	&	no	\\
HD 114783	&	$<$60	&	no	\\
HD 145675	&	$<$47	&	no	\\
HD 149661	&	$<$87	&	no	\\
HD 177830	&	$<$343	&	no	\\
HD 185144	&	$<$372	&	no    \\
HD 219134	&	$<$1334	&	no	\\
HD 220182	&	$<$60	&	no	\\
\enddata			
\end{deluxetable}

\begin{footnotesize}
\begin{deluxetable}{lcccccc}											
\tablecaption{Dust Model Results \label{dustprop}}		
\tablehead{																			
\colhead{Star}	                   &\colhead{R$_i$}	&\colhead{R$_o$}  & \colhead{T$_{dust}$} & \colhead{L$_d$/L$_*$}   &	\colhead{Dust Mass$^a$}	 & \colhead{Dust Composition$^b$} \\
 	                                     &	\colhead{AU}	&\colhead{AU}        &\colhead{K }                 &  \colhead{10$^{-5}$}       &    \colhead{10$^{-4}$ M$_\earth$} &		
	                      	}											
\startdata											
HD 10647	         &	6$\pm$0.5         & 150$\pm$150   &	70-40	     & 21.1   &   3.43       & blackbody	\\
	                   &	97$\pm$10	   &  140$\pm$50   &	33-27	     &            &		    &	silicate  - 0.25 $\micron$  or water ice \\
HD 115617	 &	96$\pm$5  	   & 195$\pm$100  &	55-45	     & 2.6     &   0.50	    & blackbody	                                              	\\
	                    &	120$\pm$20	   & 220$\pm$10   &	24-19             &            &	              &	silicate - 0.25 $\micron$          \\
HD 207129	  &	35$\pm$1	           & 45$\pm$10      &	48-44	     &   6.6    &   0.16       &	silicate - 10 $\micron$		\\
              	           &	144$\pm$1         & 200$\pm$10     &	24-12	     &             &                  &	blackbody	                                  	\\
\enddata	 
\tablenotetext{a}{The dust mass is estimated from the L$_d$/L$_*$ ratio, a dust density of 3.3 g cm$^{-3}$, a dust radius of 1 $\micron$, and a disk radius of 100 AU as defined in Jura et al. (1995).}
\tablenotetext{b}{Here, we list the assumed composition and radius (when necessary) of the grains used in the models. We assume either black body grains or silicate grains. The silicate grains
can have a radius of 0.25 $\micron$ (small grains) or 10 $\micron$ (large grains).}
\end{deluxetable}	
\end{footnotesize}

\begin{deluxetable}{lccc}							
\tablecaption{Dust Physical Properties for Stars with 160 $\micron$ Upper Limits \label{dustproplim}}		
\footnotesize					
\tablehead{							
     \colhead{Star} & \colhead{Dust Temperature$^a$} & \colhead{L$_d$/L$_*$$^b$}      & \colhead{Dust mass$^c$} \\							
                                &   \colhead{K}                                     & \colhead{10$^{-5}$}    & \colhead{ 10$^{-4}$ M$_\earth$}  						
}							
\startdata
HD 166	         &	$>$104	&	$<$0.16	& 	$<$0.33	\\
HD 17925	         &	$>$42	&	$<$0.33	&	$<$0.68	\\
HD 33262	         &	$>$27	&	$<$0.20	& 	$<$0.41	\\
HD 33636	         &	$>$34	&	$<$0.73	& 	$<$0.22	\\
HD 37394	         &	$>$23	&	$<$0.80 	& 	$<$1.51	\\
HD 52265	         &	$>$21	&	$<$2.90	& 	$<$6.02    \\
HD 69830	         &	$>$21       &	$<$1.83	& 	$<$0.90	\\
HD 72905         &	$>$36	&	$<$0.17	& 	$<$0.35	\\
HD 76151	         &	$>$25	&	$<$0.64	& 	$<$1.33	\\
HD 82943 	&	$>$43	&	$<$1.11	& 	$<$2.30	\\
HD 117176	&	$>$24	&	$<$0.50	& 	$<$1.03	\\
HD 118972	&	$>$29	&	$<$1.04	& 	$<$2.15	\\
HD 128311	&	$>$27	&	$<$1.28	& 	$<$2.66	\\
\hline							
HD 10476  	&	                  &	$<$0.19	&	$<$0.39	\\
HD 114783	&	           	&	$<$1.03	&	$<$2.13	\\
HD 13445	         &                         &	$<$0.09	&	$<$0.19	\\
HD 145675	&		         &	$<$0.33	&	$<$0.69	\\
HD 149661	&		         &	$<$0.29	&	$<$0.60	\\
HD 177830	&		         &	$<$31.8	&	$<$65.9	\\
HD 185144	&		         &	$<$0.46	&	$<$0.96	\\
HD 219134	&		         &	$<$3.51	&	$<$7.28	\\
HD 220182	&		         &	$<$0.87	&	$<$1.80	\\
\enddata	
\tablenotetext{a}{The dust temperature listed here represents the blackbody temperature 
that reproduces the ratio of the 70 $\micron$ flux and 160 $\micron$ limit.}
\tablenotetext{b}{These values  of L$_d$/L$_*$ come from Equation 2.}
\tablenotetext{c}{Upper limits on the dust mass assume the L$_d$/L$_*$ ratio from the 160 $\micron$ limit, a dust density of 3.3 g cm$^{-3}$, a dust radius of 1 $\micron$, and a disk radius of 100 AU (Jura et al. 1995).}
\end{deluxetable}

\clearpage

\clearpage
\clearpage
\clearpage
\clearpage
\clearpage
\clearpage
\clearpage
\clearpage
\clearpage

\end{document}